\documentclass[journal, article, submit, moreauthors, 10pt, a4paper]{mdpi}

\firstpage{1} 
\pdfoutput=1

\usepackage{graphicx}
\usepackage{amsmath}
\usepackage{wasysym}%
\usepackage{comment}
\usepackage{tikz, pgfplots, pgfplotstable}
\usepackage{tabularx}
\usepackage{listings}
\usepackage[colorinlistoftodos]{todonotes}
\usepackage{courier}
\pgfplotsset{compat=1.14}

\usepackage{colortbl}
\definecolor{Gray}{gray}{0.9}
\usepackage{float}

\newtheorem{theorem}{Theorem}
\usepackage{mathtools} % Bonus
\usepackage{amssymb}

\lstdefinelanguage{JavaScript}{
		  keywords={typeof, new, true, false, catch, function, return, null, catch, switch, var, if, in, while, do, else, case, break},
		  keywordstyle=\color{blue}\bfseries,
		  ndkeywords={class, export, boolean, throw, implements, import, this},
		  ndkeywordstyle=\color{darkgray}\bfseries,
		  identifierstyle=\color{black},
		  sensitive=false,
		  comment=[l]{//},
		  morecomment=[s]{/*}{*/},
		  commentstyle=\color{purple}\ttfamily,
		  stringstyle=\color{red}\ttfamily,
		  morestring=[b]',
		  morestring=[b]"
		}
 \theoremstyle{mdpi}
 \newcounter{thm}
 \newcounter{ex}
 \newcounter{re}
 \theoremstyle{mdpidefinition}

\Title{Mayall: A Framework for Desktop JavaScript Auditing and Post-Exploitation Analysis}

\Author{Adam Rapley, Xavier Bellekens, Lynsay A. Shepherd, and Colin McLean}
\AuthorNames{Adam Rapley, Xavier Bellekens, Lynsay A. Shepherd, and Colin McLean}

\address[1]{%
School of Design and Informatics, Abertay University, Dundee, DD1 1HG, United Kingdom  \\Email: x.bellekens@abertay.ac.uk; lynsay.shepherd@abertay.ac.uk; c.mclean@abertay.ac.uk}

\corres{Correspondence: x.bellekens@abertay.ac.uk; Tel.: +441382808482}

\abstract{Writing desktop applications in JavaScript offers developers the opportunity to write cross-platform applications with cutting edge capabilities. However in doing so, they are potentially submitting their code to a number of unsanctioned modifications from malicious actors. Electron is one such JavaScript application framework which facilitates this multi-platform out-the-box paradigm and is based upon the Node.js JavaScript runtime --- an increasingly popular server-side technology. In bringing this technology to the client-side environment, previously unrealized risks are exposed to users due to the powerful system programming interface that Node.js exposes. In a concerted effort to highlight previously unexposed risks in these rapidly expanding frameworks, this paper presents the Mayall Framework, an extensible toolkit aimed at JavaScript security auditing and post-exploitation analysis. The paper also exposes fifteen highly popular Electron applications and demonstrates that two thirds of applications were found to be using known vulnerable elements with high CVSS scores. Moreover, this paper discloses a wide-reaching and overlooked vulnerability within the Electron Framework which is a direct byproduct of shipping the runtime unaltered with each application, allowing  malicious actors to modify source code and inject covert malware inside verified and signed applications without restriction. Finally, a number of injection vectors are explored and appropriate remediations are proposed.}


\begin{document}

%%%%%%%%%%%%%%%%%%%%%%%%%%%%%%%%%%%%%%%%%%
%% Sections that are not mandatory are listed as such. The section titles given are for Articles. Review papers and other article types have a more flexible structure. 

%% Only for the journal Gels: Please place the Experimental Section after the Conclusions

%%%%%%%%%%%%%%%%%%%%%%%%%%%%%%%%%%%%%%%%%%
\section{Introduction}\label{intro}
Application designers often seek to produce high quality, cross platform applications in order to maximize their customer base. This introduces a number of difficulties regarding code reuse, with applications having to be modified, re-written or recompiled for each individual operating system.

Enter the explosion of new, fashionable languages and their counterpart interpreters and compilers. Such \textit{`write once, run everywhere'} languages, with their multi-platform out-the-box paradigm, attempt to revolutionize the way in which native applications are developed. These languages, such as Go and Node.js, have garnered rapid popularity through their inherent power and ease of development \cite{nodeRapidPopular}.
In the case of Node.js, this rapid rate of expansion brought with it a staggering number of modules and frameworks, along with an ever growing list of security issues.

Electron is one such framework aiming to bring this JavaScript runtime to the native operating system, unifying web technologies and distilling them into what is effectively a native executable. In doing so, developers are not only potentially opening themselves up to traditional web exploits such as Single-Origin Policy misconfiguration \cite{meyerovich2010secure} --- albeit this time in the much more dangerous context of the operating system --- but also exploits that are unique to the framework in question.

Frameworks such as Electron, NW.js and to some extent, the Chrome Embedded Framework, are rapidly expanding and as such, it is important to address the security surrounding these products. There has been an increasing number of recent discoveries and events which bring into question the level of focus on the security surrounding the modules available on \textit{npm}, the de facto standard for Node.js package management. As Electron is beginning to emerge as the clear winner of the two Node.js based frameworks, this paper will focus predominantly on the security surrounding Electron and the increasingly popular applications which are built upon it.

When contemporary frameworks grow at such exponential rates, including adoption from major players such as Microsoft \cite{microsoftUseNode}, NVIDIA \cite{NVIDIANode} and Slack \cite{appsOnElectron}, it is crucial that the security awareness surrounding the product does not go amiss. This reliance on the Node.js JavaScript runtime and the package manager that is used so prolifically with it compounds the security issues, as it adds additional levels of complexity which each require individual consideration with regards to security. There is comparatively little in the way of security research for Node.js and its accompanying frameworks when set side by side with other native programming languages, and what has been implemented in the way of security fixes is often reactive, rather than proactive. This is especially true with regards to the debacle over the \textit{kik} module in March 2016, where an author unpublished a package which many other packages depended on \cite{kik}.

This crucial oversight, highlights a clear need to analyze the platform for other potential issues, and to consider remediations for them before they occur. By investigating the method by which dependencies are handled and the way in which code is executed, the Mayall framework \footnote{Named after the astronomical body, Mayall's Object, which consists of two colliding galaxies, the framework's primary aim is to inject a module into the original application and modify its execution.} presented in this paper allows developers to do just that.  This paper explores the current state of security surrounding the Electron framework and the underlying Node.js constructs and demonstrates real world risks associated with this through the introduction of a post-exploitation framework built in Node.js, leveraging the modular structure of the language. The contributions from this paper are threefold:

\begin{enumerate}
	\item An exploit which can take advantage of popular Electron based applications such Slack, Discord, etc.
    \item An investigation of issues inherent of the Electron Framework.
    \item An Open Source Extensible Security framework for analyzing Node.js applications.
\end{enumerate}

The remainder of the paper is organized as follows: In section \ref{litreview}, an assessment of the current state of JavaScript security is performed and categories of vulnerabilities, language constructs and application management are discussed. This current review is taken further in section \ref{methodology} where we expose the vulnerabilities of fifteen popular applications built with Electron, moreover we present a proof of concept exploit which allows us to take advantage of Electron based applications. The results from the audit are presented in section \ref{results} and a discussion on how to remediate the risks from exploitation is proposed in section \ref{discussion}.

%Background
\section{Background and related work} \label{litreview}
JavaScript security is a long discussed and multi-faceted affair; this is no different within the context of the Electron and Node.js runtimes. Now that the JavaScript language can be run natively, outwith the context of the browser, additional security concerns are brought to the fore. This is compounded by the rapid growth experienced by both JavaScript and Node.js over the past few years; JavaScript has consistently been top of the most popular technologies in the Developer Survey Results produced by \cite{soDSR}.

This section explores the current security themes that are directly affecting the Node.js landscape. Specific examples are drawn upon where applicable to highlight the relevance of these themes when applied to Node.js. Whilst on the surface it may appear that some of these examples are self contained and acutely specific, the general risks discussed can be applied to Node.js and its desktop application counterpart, Electron. It is therefore important to discuss these precursor attacks that have direct relevance to these new languages and frameworks going forward.

\subsection{Web based JavaScript} \label{webBasedJavaScript}
When looking at traditional web-only based applications written in JavaScript, there are a series of recurring patterns that introduce a possibility for security vulnerabilities. For example, contained within JavaScript are a set of functions which allow for dynamic parsing and production of JavaScript code, the primary of which is \texttt{eval()}. One particular study by Richards~textit{et~al.} took a large sample of the top 10,000 websites and discovered that 89\% of them used JavaScript. Furthermore, they discovered that \textit{``over 82\% of the top 100 pages use \texttt{eval}, and 50\% of the remaining 10,000 pages do as well''} \cite{evalMenDo}.
The primary issue that occurs with the use of~\texttt{eval} is when arbitrary and unsanitized user data is passed to the function. When this occurs, it is possible for an attacker to execute the string they provide to the web application as JavaScript code. Whilst static code analysis often highlights security issues within an application, this becomes more challenging when dynamic elements such as \texttt{eval} are introduced. One method of analyzing such code is to perform an automatic transformation of \textit{``many common uses of~\texttt{eval} into other language constructs with the use of a dataflow analyzer''}~\cite{evalMenDoFix}. These transformations are able to make some headway towards restoring the effectiveness of static code analysis on JavaScript applications. This is as a direct result of reducing the number of malleable and untrusted variables within the code, which subsequently reduces the number of mechanisms available to exploit such code.

While~\texttt{eval()} is arguably the most relevant security risk when applied to modern cross-platform JavaScript applications due to it's alarming prevalence in Node.js, there still remain a large number of additional JavaScript vulnerabilities that apply to both web and desktop JavaScript, the most common of which is cross-site scripting (XSS)~\cite{owaspTopTen}. Marked as third in the OWASP Top 10 vulnerabilities list, XSS works similarly to injection vulnerabilities (such as those described above) in that it arises from a mishandling of user input and can result in arbitrary JavaScript execution within the web application~\cite{binnie2016x}.  Furthermore, JavaScript is used in the development of browser add-ons i.e. Web Extensions in Mozilla Firefox.  Such vulnerabilities have the potential to allow for the development of malicious extensions, posing a security risk \cite{shepherd2014reducing}.

\subsection{Node.js and Electron} \label{NodeAndElectron}
\subsubsection{Node.js} Node.js is a JavaScript runtime that is built upon the V8 JavaScript engine developed by Google. Its primary aim is to provide a highly scalable environment on which JavaScript code can be run, allowing it to be deployed on desktop and server-side environments. Its focus is driven towards asynchronicity and concurrency in handling a large volume of connections whilst preventing thread-locks within the application \cite{aboutNode}.  It can be reasoned that Node.js is more akin to a standard library than a web framework and is comparable to PHP, Ruby or Python rather than the frameworks that run atop these platforms such as Laravel, Ruby on Rails and Flask. This is an idea bolstered by Eloff~\textit{et~al.} and his analysis of Node.js as a baseline platform for web services \cite{eloff}.

Its adoption by commercial players such as Uber \cite{uberUseNodeFile} and Netflix, \cite{netflixUseNode} amongst others has helped to springboard the language towards success.
%\begin{comment}A big part of that success has also been down to the mass adoption of the language by developers, with Node.js having the fastest growing package management library of any modern language \cite{modulecounts}.\end{comment}%

As it aims to be a powerful programming language in its own right, this requires it to contain a number of libraries and Application Programming Interfaces (APIs) that allow it to perform \textit{``low level networking, basic HTTP server functionality, file system operations, compression and many other common tasks''} \cite{ojamaa2012security}\cite{assSec}. As a result of this, the \texttt{exec()} command was introduced to the Node.js specification in order to expose shell interaction functionality on the operating system level. Similar to the way \texttt{eval()} works, data that is provided to \texttt{exec()} is interpreted as a command, however it is passed to the operating system shell rather than the JavaScript engine \cite{synode}.

\subsubsection{Electron}
The Electron framework is an open source library designed for cross-platform application development in web based languages HTML, CSS and JavaScript \cite{aboutElectron}. It combines Chromium and Node.js into a single package so that traditional back-end and front-end code can be run within a single runtime environment.

A basic Electron application consists of three components: the \texttt{package.json} file, the Electron executable binary and the JavaScript source code.

\subsection{Node Security}
\label{NodeSec}
The tendency for writing insecure \textit{eval} statements is not a scenario that is just limited to the web --- desktop JavaScript is also affected by this \cite{synode}. As we see an increase in desktop JavaScript applications being shipped (examples include the increasingly popular Slack and Discord \cite{appsOnElectron}), the programming paradigms that were popular on the web, are also unsurprisingly popular on desktop frameworks.
The transference of such insecure coding practices are dangerously widespread as  Staicu~\textit{et~al.} show~\cite{synode}. The study was carried out across a sample set of 235,850 Node.js modules. Their results demonstrate that approximately 20\% of these modules made use of the potentially unsafe \texttt{eval()} and \texttt{exec()} API calls directly or indirectly after all the first and second level dependencies were accounted for. A reasonable majority of these modules did not directly call these APIs however, with only 3\% and 4\% of the sample calling the \texttt{exec()} and \texttt{eval()} APIs directly --- the remainder were run within the module dependent code~\cite{synode}.

Whilst popular belief may lead to the conclusion that these insecure modules are unpopular, this hypothesis is challenged by the discovery that the contrary is true and that in fact \textit{``various vulnerable modules and injection modules are highly popular, exposing millions of users to the risk of injections''} \cite{synode}.

The immediacy of these findings does not present itself until it is noted that the sandboxes present in the web browsing environments (such as Chrome and Firefox) are not present within the Electron framework \cite{noElectronSandbox} and that any application running on Node.js has full system access equivalent to the account that is running the process.

\begin{comment}
The idea that Node.js is a language designed around small modular imports is something that was discussed during a phone call with the founder of \^{}liftsecurity and the Node Security Platform, Adam Baldwin. He posited that Node is so powerful in part due to its wealth of developers who contribute `single function modules' that aim to do one job very well. However, one of the downsides to decentralized code repositories is the lack of accountability that comes with it. This idea is backed up by research into remote inclusions which put forward the important notion that \textit{``whenever developers of a web application decide to include a library from a third-party provider, they allow the latter to execute code with the same level of privilege as their own code''} \cite{whatYouInclude}.
\end{comment}

\subsection{The npm Registry and Module Security} \label{npmLit}
When the number of external \textit{includes} present in an application increases, so does the potential for unchecked vulnerabilities and the level of trust placed in unknown module authors. This idea is backed up by research into remote inclusions which put forward the important notion that \textit{``whenever developers of a web application decide to include a library from a third-party provider, they allow the latter to execute code with the same level of privilege as their own code''} ~\cite{whatYouInclude}. The package management system that is used for Node.js is known as \texttt{npm} and is where the vast majority of Node modules reside. Additionally, as part of the module install process, npm exposes the functionality to run additional scripts on the target machine \cite{npmScripts}. These so-called \texttt{preinstall} and \texttt{postinstall} scripts have historically been a cause for concern within the npm registry and Node.js landscape. These script hooks introduce a new danger to the npm and Node.js environment as the scripts will run on the host with the same privileges as the user installing the module.

\subsubsection{Typosquatting}
Tschacher~\textit{et~al.} brings to light the dangers of typosquatting within package managers in his dissertation he explores the malicious module that was released into the npm registry by a malware author on 26th January, 2015 \cite{typosquat}\cite{mcluskie2018x}. The module in question was known as \texttt{rimrafall} and contained \textit{``a preinstall hook that executed the command \texttt{rm -rf /*}''} \cite{malModNpm}\cite{ojamaa2012security}. As a result of this pre-install hook, any unsuspecting user that entered \texttt{npm install rimrafall} in a terminal prompt would end up with all their files being deleted from the machine. Whilst this module was removed from the registry less than two hours after being published, it does highlight the risks present in blindly \textit{requiring}\footnote{Similar to a Python \texttt{import} statement, a \texttt{require} statement is the equivalent method by which external JavaScript modules are imported to a JavaScript application.} numerous modules into a project. The primary goal of this module appeared to be to highlight that ``\texttt{npm install} can be as dangerous as \texttt{curl dangerous.com | sh}'' \cite{rimrafall}.

Whilst Lift Security provide tools to actively monitor modules that are \textit{required} as well as systematically audit the most popular modules present on npm \cite{malModNpm}, the fact still exists that malicious modules can be introduced to the npm registry. Although methods exist to prevent scripts from running on install, the majority of users simply do not apply these checks when \textit{requiring} modules into their codebase \cite{typosquat}.

This lack of concern and consideration employed by some developers is evidenced in research performed on the effectiveness of typosquatting in a wide range of package managers for a multitude of different languages. 214 different packages were uploaded to the respective registries with typos in their names, each of which acquired 92 installations on average per package \cite{typosquat}. Whilst this may not sound like much, \textit{``the most installed package (\texttt{urllib2}) received 3929 unique installations in almost 2 weeks''} and \textit{``the most installed package per day was \texttt{bs4} with 366 unique daily installations on average''} \cite{typosquat}.

What is more alarming is the origin where these downloads occurred, as well as the prevalence of installs that were run with administrative privileges. It was found that 43.6\% of module downloads and inclusions were run with administrative rights \cite{typosquat}, thereby giving the accompanying install hooks administrative access on the machines where they were installed. In addition to the checking of administrative rights, \cite{typosquat} performed reverse DNS lookups with the installed module and discovered that his module had been installed on a surprising number of government and educational domains.

\subsubsection{Trojan Modules}
In addition to the direct installs of first level dependencies and the evident risks associated with directly installing unchecked code, there exists a risk of hidden malware further down the dependency tree. If a malicious author can influence code that is included in a numerous amount of other modules, the infection rate will be vastly increased. This exact scenario has occurred in the past on a variety of modules, however \cite{whatYouInclude} points to a specific instance with popular jQuery plugin, \texttt{qTip2}. Nikiforakis~\textit{et~al.} state that ``The \texttt{qTip2} library was modified, and the malicious version was distributed for 33 days''~\cite{whatYouInclude}. The compromised code made frequent callbacks to a specific IP address located in Russia and transmitted data including \textit{``[the] site's hostname, [the browser's] user agent, and the referer''} \cite{qTip2GitHub}.

These types of risk are not exclusive to WordPress and the web environment, this threat has presented itself in the npm registry. An npm developer encountered legal issues with Kik Interactive Inc. --- developers of Kik Messenger --- when he published a module under the name `kik'. The npm developer was unable to come to an agreement with the company and after a fall out with the npm registry over the use of `kik' as his module name, the developer took the decision to remove all 273 packages that he authored from the registry. As a result of this, thousands of Node.js applications --- including the widely used \texttt{babel}\footnote{Babel (https://babeljs.io/) is a JavaScript compiler used by Facebook, CloudFlare and Netflix, amongst others.} library --- started issuing dependency errors and failed to execute \cite{kikRegister}.

In addition to the breaking changes that were made due to the unpublishing from the npm registry, the removal of these modules presented an even greater threat. Every single module name-space that had been removed had now become available for any developer to claim control over. This allowed any actor, either benevolent or malicious, to potentially inject code into thousands of Node projects \textit{requiring} any of the widely used modules that he had authored \cite{kik}. npm responded reflexively by blocking the registration of any of the removed modules as well as applying the same tactic to any future removed application, but not before a vast number of modules dropped from the registry had already been claimed by other developers.

\subsection{Developer Attitudes} \label{devAttLit}
As with a lot of systems, updating is often an easy way to ensure that applications are free from vulnerabilities. However this is a major issue with the npm repository. As a result of the codebase being highly distributed over an exceedingly large number of developers, high code quality and a guarantee of constant security updates is not always available.
Whilst auditing the top npm packages, \cite{synode} discovered a number of vulnerabilities within modules which they responsibly disclosed to the developers. They stated that \textit{``most of the developers acknowledge the problem, and they want to fix it. However, over the course of several months, only three of the 20 vulnerabilities have been completely fixed. The majority of issues are still pending, showing the lack of maintenance for many of the modules involved.''}.
This lack of maintained code is not the only issue present in the Electron and Node.js landscape; the prevalence of version pinning also prevents a large majority of applications from being updated even after vulnerabilities are patched. This has become the mainstream method of dependency management after GitHub actively recommended \textit{``[setting] a fixed version number (\texttt{1.1.0} instead of \texttt{\^{}1.1.0}) to ensure that all upgrades of Electron are a manual operation made by the developer''} \cite{electronUpdate}. This prevents applications suffering when breaking changes are introduced into the Electron framework, however it does place the onus on the developers to pay attention to updates and manually push these security fixes to the end users.

The following section expands on the work done by previous researchers,  with a sharper focus on desktop applications written in JavaScript and JavaScript based frameworks, by presenting a framework centered around the security of Node.js and it's counterpart front-end, Electron. The framework is highly extensible and allows for the use of cross-platform payloads for remote execution of malicious code, taking advantage of the lack of code signing enforcement either remotely or locally.

\section{Design and Implementation}\label{methodology}
This section aims to highlight the stages of development of the framework. The framework provides two algorithms presented in subsections \ref{appScanner.js} and \ref{nspCheck.js} respectively. Both aim at assisting a security analyst during an audit with automated JavaScript vulnerability scanning. Additional security vulnerabilities are brought to the fore in \autoref{updateProcess} with regards to how Electron handles updating application bundles and an exploit is demonstrated in \autoref{malUpdate}. This notion of exploit development is expanded further with the production of a fully functional `backdoor module' in \autoref{localInject} which discusses the methods by which Node.js malware can be introduced into a system and is rounded out in \autoref{toolkit} with the amalgamation of the produced tools into one cohesive framework that can be used for JavaScript security analysis.

\subsection{Application Auditing} \label{auditing}
Due to the high prevalence of insecure modules present within the npm repository, it is deemed appropriate to perform an analysis of modules in use by some of the most popular Electron based applications. Electron applications are bundled into what is known as an \textit{asar archive}, which is a concatenation of all files in the source code folder, similar to a tar archive. Electron can then access files from this archive during execution of the application. The \texttt{asar} package can be installed globally onto a system by using \texttt{npm} and the command \texttt{npm install -g asar}, after which point the \texttt{asar} command can be called directly from a terminal prompt.
The asar archiving process does not contain any encryption or obfuscation and the tool is freely available and open source. As a result of this, it is possible to issue the command \texttt{asar e app.asar app} to retrieve the entire source code folder for an application, complete with formatting, comments and module dependencies as written by the developer. It is the latter which is of interest during this phase of the investigation as attempts will be made to discover if any of the modules contain vulnerabilities.
Modules that have been \textit{required} by either the application author directly (henceforth referred to as \textit{primary modules}) or any of its dependencies (henceforth referred to as \textit{tertiary modules}) are located in the \verb|node_modules| folder within the decompressed asar archive. Each of these modules contains an individual \texttt{package.json} file which is specific to that module. This file contains details such as the name, version and author of the module, but also additional information such as the repository url for that module. Of the modules surveyed, only a negligible amount of modules did not provide a GitHub repository location and it was therefore very easy to further analyze these modules through commits and issues present on the respective repositories.

\subsection{appScanner.js} \label{appScanner.js}
An algorithm was developed to assist in the version analysis of all imported modules, as the number of tertiary modules can be vastly larger than the number of primary modules, thereby creating excessive amounts of workload for a manual audit, as shown in \autoref{results}.
By interfacing with the GitHub API, it was possible to pinpoint the individual git commit which corresponded to the release number inside a module's \texttt{package.json} file. This was done by querying the repository listed within the \texttt{package.json} file for tagged commits with the version number of the module. Following this, a comparison of the returned commit to the latest commit on the master branch was made and a count was returned of the number of commits the master branch was ahead of the imported module.
Modules that were a substantial number of commits behind the master branch --- above or equal to 150 --- were considered to be suspect packages and could help an analyst look in the right places for existing bugs and vulnerabilities which would affect an application.

Although being a large number of commits behind the master branch may at first seem alarming, a lack of pulled commits may themselves not necessarily indicate a directly actionable exploit or vulnerability --- this idea is discussed in more detail later in \autoref{appScannerDiscussion}. If an application has not been updated to the latest build, it may be to prevent breaking changes rather than a lack of concern for updating.
\subsection{nspCheck.js} \label{nspCheck.js}
While checking how far a module is behind the master branch helps to give indicators to the security status of an application, a git commit record does not immediately indicate security flaws within the respective modules without additional  analysis. To this end, another algorithm was developed allowing immediate highlighting of known and reported security vulnerabilities within a module.

The Node Security Platform maintains a list of security advisories along with a corresponding API. By comparing the installed module versions against this NSP database, immediate security vulnerabilities can be flagged to developers. By querying this API and parsing the response data into an easily human readable format, the algorithm makes strides towards building a toolkit for security researchers and analysts assessing overall Node.js security within a network. 

\begin{figure}[h]
\centering
\includegraphics[width=88mm]{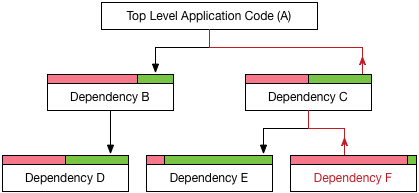}
\caption{Nested dependencies}
\label{fig:depGraph}
\end{figure}

When auditing for vulnerabilities, it does not suffice to simply check the \texttt{package.json} for the main application. Although a developer may \textit{require} only modules that are up-to-date, they are not directly responsible for the management of multi-level dependencies, with this responsibility falling to the module author. As a result of this, vulnerabilities can be introduced despite a developer including only up to date version numbers. This is demonstrated in \autoref{maths:depProof} and \autoref{fig:depGraph}, where it can be seen that whilst the developer has required up-to-date module versions, those modules in turn could depend on vulnerable code which is subsequently imported into the application.
\begin{theorem} \label{maths:depProof}
\renewcommand{\qedsymbol}{$\blacksquare$}
Let $A$ be the main source code, where $B,C = 1^{st}$ are level dependencies and $D,E,F = 2^{nd}$ are level dependencies. Let $X_n$ be the module version, where $n=1$ indicates the most recent version and $n=0$ represent outdated versions. Let $X_r$ be the set of requirements for a module.
  \begin{gather*}
    A \supset \{B,C\} \supset \{D,E,F\} \\
    A_r \mapsto B_1 \cup C_1 \\
    B_r \mapsto D_1 \\
    C_r \mapsto E_1 \cup F_0 \\
    \therefore A \mapsto F_0 \qedhere
  \end{gather*}
 This demonstrates that a single pinned version requirement further down a dependency tree can result in the inclusion of outdated modules, despite the developer updating all of their primary dependencies.
\end{theorem}

Figure~\ref{fig:depGraph} illustrates the dependency tree of an application, and demonstrates how outdated modules can be included into top level applications. Nested dependencies can introduce vulnerabilities through version pinning. Arrows represent dependency calls, while the red and green sections represent vulnerable and patched module versions respectively. Although an alarming percentage of the applications tested did contain known vulnerable elements (\autoref{tab:nspChecker}), there still were a number of large name, well known applications which passed this scan with no vulnerabilities being detected. With this in mind, it was necessary to widen the scope to audit other aspects of the application processes, one of which was updating.

			\begin{table}[ht] % NSP Scan
			\centering
				\begin{tabular}{ccc}
					\hline
					Application & Total N\textsuperscript{\underline{o}} Advisories & Highest CVSS Score \\
					\hline
					1Clipboard & 25 & 7.5 \\
					Atom & 25 & 7.5 \\
				    Caret & 2 & 7.5 \\
					Discord & 3 & 7.5 \\
					Ghost & 0 & --- \\
					Hyper & 0 & --- \\
					Mattermost & 0 & --- \\
					MongoDBCompass & 17 & 7.5 \\
					NVIDIA & 11 & 7.5 \\
					Popcorn Time\footnote{Popcorn Time is not an Electron app and is instead based on NW.js.} & 39 & 7.5 \\
					Popkey & 7 & 7.5 \\
					Slack & 0 & --- \\
					Tofino Browser & 0 & --- \\
					SteelSeries & 1 & 7.5 \\
					Wire Messenger & 5 & 7.5 \\
					\hline
				\end{tabular}
				\caption{An automated scan of the applications against the Node Security Platform advisories list. The number of advisories may contain duplicate entries if the module has been \textit{required} more than once.}
				\label{tab:nspChecker}
				\end{table}

\subsection{Malicious Update Process} \label{updateProcess}
Updating Electron applications is typically managed via the Squirrel Framework which is provided to Electron developers as an interface through the \texttt{autoUpdater} module shipped with Electron. The update process is not uniform across platforms with the Windows process requiring additional steps in order to produce the update packages.

\subsubsection{Update Electron Packages} \label{buildUpdate}
When Electron applications are updated on macOS, the only step that is taken is to recompile the entire application and push it to an update server. This update server is then polled at application startup by Electron applications running the \texttt{autoUpdater} code. For Windows, it is possible to create \textit{delta packages} which contain the changes made since the last version.

The update server follows a specific directory layout where the platform specific updates are stored in sub-folders for that platform. The Windows subdirectory contains a \texttt{RELEASES} file which contains a list of packages available on the server in the format `\verb|SHA1_checksum| \verb|filename| \verb|size_in_bytes|'. This \texttt{RELEASES} file is queried by the remote applications at startup to determine if an update is available. If one of the entries in the \texttt{RELEASES} file contains an application with a greater version number than the one provided by the application, the update process will automatically download and update to the latest version without any user input required.

Automatic updates on Windows will occur so long as code signing has not previously been put in place and the version number is higher than the current version performing the update check. On the other hand, macOS applications will not automatically update unless the update package is signed with a valid code signature. For this reason, a proof of concept was built on the Windows platform to eliminate additional overheads which are outwith the scope of this study, namely code signature bypasses.

\subsubsection{Enumerating Update Endpoints}
As a result of requiring no user input, it may be possible to covertly install malicious updates without user interaction. In order to demonstrate the vulnerability, a freely available application built on the Electron Framework was downloaded and it's contents extracted with \texttt{asar e app.asar app}. The code in the resultant \texttt{app} folder was then examined and additional code was inserted. The added code resulted in the writing of a new file to the user's home folder on launch.

Due to the lack of code signing in effect on the application, and the lack of a secure HTTPS connection to the update server, it was possible to man-in-the-middle (MITM) the entire update process. When the application was launched, the connection request to the remote update server was intercepted and handled by a rogue update server that mimicked the responses that the application was expecting.

In order to build this rogue server, the update endpoints needed to be determined. This was achieved by setting an HTTP server handling all requests from the application. By modifying the hosts file to redirect the target application to localhost, all URL endpoints could be enumerated by the rogue server. It was subsequently possible to recreate the folder structure required for a Windows update and have the rogue server resolve the requests made by the insecure application.

\subsubsection{Building Malicious Updates}
\label{malUpdate}
The Windows update process involves downloading a delta package (a package containing the changes made since the last version) or, if one is unavailable, a complete package from the server. The package system in use is based on the Windows NuGet Package Manager and application updates are provided to the Squirrel Framework as \texttt{.nupkg} files.

A \texttt{.nupkg} file is compressed in the same way as a \texttt{.zip} folder and contains the source code for the application along with associated metadata such as version numbers. By unzipping the full nupkg file for the most recent version of the application and modifying the metadata and source code, a fully functional update could be created and served by the rogue server.

As a result of these steps, the application automatically updated on launch to the `most recent version' as supplied to it by the rogue server, thereby writing a file to the user's home folder. As a presumed formality, the application presented the user with a dialog box informing that an update was available and asked the user if they would like to update now or later. However, this dialog box only appeared after the automatic update had already been installed and it was irrelevant whether the user clicked `Install' or `Later'. If the user clicked `Install', the application would open a web page in the user's default browser showing the patch notes for the version just installed. To alleviate suspicion from a user, this webpage (\url{http://1clipboard.io/update}) was cloned and served to the user through the malicious server.

    \begin{lstlisting}[caption=Mayall Interaction Pseudocode - Target Code, label=code:inj]
const inj = new Function(conn_port) {
  net.createServer(function (socket) {
    // On the receipt of 'exec' data, execute as a commend.
    socket.on('exec', function (data) {
      exec(command, (err, stdout, stderr) => {
        socket.write(stdout);
      });
    });
  }).listen(conn_port);
}
\end{lstlisting}
%\subsection{Locally Effecting Change in an Application}
\subsection{Malicious Exploit Design} \label{localInject}
While injecting via an update is a viable infection method if the application does not correctly implement code signing, there remain a number of applications which do code sign and thus this method may sometimes be impractical. Despite the fact that code signing is in place during the update process, this security check is not enforced at application runtime and code signing is disregarded. To this end, if it is possible to gain access to a machine --- through a social engineering or phishing attack for example --- it would be possible to modify the application source code to inject malicious content. This is possible as the Electron executable does not check the code integrity of the application on runtime.

\begin{lstlisting}[caption=Mayall Interaction Pseudocode - C2 code, label=code:c2]
const c2 = new Function(local_port) {
  io.on('connection', function(socket) {
    // Initial agent activation
    dbInterface.addToAgents(socket.id);
    socket.on('hello', function(msg) {
      socket.emit('c2','world');
    });
    // Returning a response from a command exection
    socket.on('agentChatter', function(msg) {
      process.stdout.write(msg)
    });
  });
  // Commands can be sent to specific agents with specific types
  // Type classification can be seen in line 5 of listing 1.2
  exports.pushCmd = function(agent, type, command) {
    io.to(agent).emit(type, command);
  }
}
\end{lstlisting}

\subsubsection{Malicious Exploit Design - A Dropper Module}
Before the injection of an application is demonstrated, it is  necessary to generate a payload that is able to communicate with and receive commands from a remote Command and Control (C2) server as demonstrated in \autoref{code:inj} representing the target code, and \autoref{code:c2} representing the pseudo code for the C2 Server. As Node.js is the underlying language that Electron is based on, the basic module was built in Node.js without Electron in mind --- Electron is primarily used as the visual bolt-on to a Node.js application, exposing the Graphical User Interface (GUI) and other interaction methods to an end-user and Node modules can still be run underneath Electron.

In order to receive commands, a websocket functionality was introduced, opening a port on the target machine on which \textit{mayall.js} listen for commands. This can be seen where the payload waits for data tagged with the string 'exec', the contents of which it executes in an \texttt{exec} command, writing the results back out to the socket. Once the module is running on a target machine, it is then possible to connect with a networking tool such as netcat.

\subsubsection{Covertly Embedding Modules}
Running a node module outright from the command line, although effective in demonstrating Node.js capabilities, is not going to have a high conversion rate from payloads delivered to remote shells returned. In order to have a greater activation rate from the Mayall payload, it was deemed necessary to embed it within applications that a user trusts and installed themselves. An additional benefit of embedding an application with the Mayall malware (rather than running it directly) is that any firewall notifications will be requested on behalf of that injected application. For instance, if Mayall is injected into Slack --- a popular team communication product --- any initial firewall request by Mayall to open ports will appear to come from Slack and not the malicious module. Furthermore, once this request has been accepted once, it will continue to be allowed to run on future execution of Slack.

\begin{figure}
\centering
	\includegraphics[width=88mm]{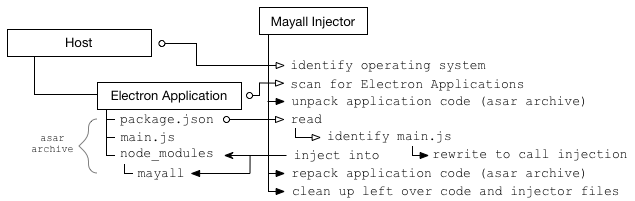}
\caption{Flow of execution for injecting malicious modules into trusted Electron Applications.}
\label{fig:nodemodules}
\end{figure}

The mayall.js payload is developed further to allow agnostic infection across multiple operating systems and applications. A list of popular Electron applications is compiled and shipped alongside the module, complete with an automatic injector. \autoref{fig:nodemodules} demonstrates the method by which the payload embeds itself within an application, starting with an initial operating system detection. After determining the platform it is ran inside (exposed through Node.js' \texttt{process.platform} variable) the injector is subsequently handed off to the appropriate dropper module. This module will search the typical installation locations for Electron applications for the system type and inject mayall malware into any applications that it discovered.

This is achieved by first extracting the asar archive associated with the application and executing a read of the \texttt{.main} field from the \texttt{package.json} file located within the archive. This field details the entry point of the application. If a valid file is discovered, the dropper downloaded the mayall module into the \verb|node_modules| folder in the extracted archive. This is followed by a pre-pending of the entry file with a require statement for the newly downloaded module, causing it to be loaded into memory when the application next executes. Once in place, the injector then repackages the asar archive and cleans up any files that were left behind as a result of the execution.

\begin{figure*}
\centering
\includegraphics[width=5in]{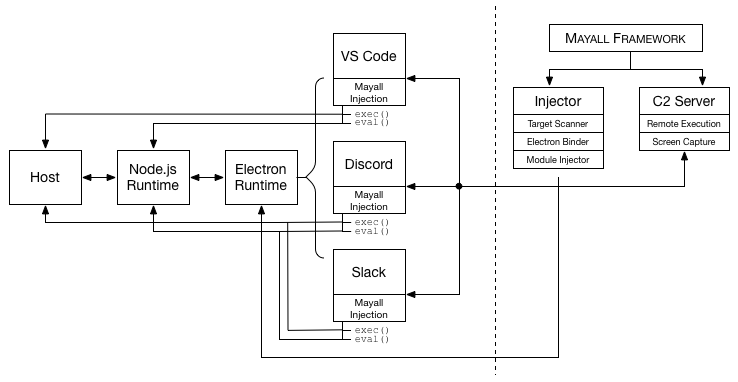}
\caption{Framework flow between the infected system and command and control server running under Mayall.}
\label{fig:mayallFlow}
\end{figure*}

As with many modern day applications, developers of Electron applications exhibit a tendency to have their applications launch on boot. As a result of this software design trait, persistence is gained when Mayall is injected into an application which exhibits this behavior --- examples include the highly popular Slack, Discord and Tidal Music applications (\autoref{fig:mayallFlow}).

\subsubsection{Executing JavaScript Natively on a Target Machine}
As this injection module is written in Node.js, it may be assumed that the Node.js is a necessary installation prerequisite for executing a successful attack. However this is not the case. If Electron applications are the target then it is possible to use operating specific scripts (for example, batch files on Windows or bash scripts on *nix systems) to identify these applications. Once identified, the injector module can be executed with the Node.js binaries that are provided alongside the Electron executables. As aforementioned, the Electron application does not concern itself with the integrity of the code that is being run and as such it would be possible to decouple the main application code from the Electron application and instead run the malicious code against this decoupled binary including a cleanup and replacement of the old code base.
The Node.js binary can often be found in unexpected locations. NVIDIA is one such example. When installing the appropriate drivers for an NVIDIA graphics card, a web helper is also installed for the NVIDIA GeForce Experience companion application. This applications handles driver updates, game optimisation and the ability to record and stream gameplay through ShadowPlay. Additionally, this application is installed by default when drivers are installed for the graphics card.

Contained within this package is a binary titled \texttt{NVIDIA Web Helper.exe} which upon further inspection is a Node.js executable which has been resigned by NVIDIA. This provides a binary upon which malicious code can be run with the added bonus of being signed by NVIDIA, therein bypassing certain security features provided by Windows --- namely application white-listing \cite{NVIDIANode}.

\subsection{A Framework for Post-Exploitation} \label{toolkit}

In a concerted effort to raise awareness and kickstart the introduction of actionable change in the JavaScript security landscape, the Mayall Framework will be released as an open source tool for the open source community, allowing security researchers to efficiently and effectively drill down on the security issues plaguing modern JavaScript applications, in a similar vein to other frameworks such as MetaSploit and PowerShell Empire.

\section{Results} 
\label{results} 
This section presents the findings of the investigation carried out on the different applications.
\vspace*{-4mm}
\subsection{Scanning Module}
The scanning module produced in \autoref{appScanner.js} was run against a number of the most popular applications written in the Electron framework, the results of which can be seen in \autoref{tab:commitsBehind}.

This shows how many modules are \textit{included} directly by the developer, followed by the subsequent total number of modules that are imported to the application through the full dependency tree. The commits column shows the number of commits behind the master branches in the associated GitHub repositories, as well as the number of commits each module is behind on average. 
			\begin{table}[ht] % GitHub Scan
			\centering
				\begin{tabularx}{108mm}{ccccc}
					\hline
					Application & N\textsuperscript{\underline{o}} inc. & N\textsuperscript{\underline{o}} deps. & Commits & Avg. comm. \\
					\hline
					1Clipboard & 24 & 374 & 48724 & 157.23 \\
					Atom & 56 & 578 & 34181 & 82.36 \\
					Caret & 29 & 89 & 2615 & 46.70 \\
					Discord & 10 & 130 & 9805 & 89.14 \\
					Ghost & 10 & 45 & 1014 & 25.35 \\
					Hyper & 17 & 69 & 2039 & 29.99 \\
					Mattermost & 11 & 85 & 7652 & 99.38 \\
					MongoDBCompass & 86 & 841 & 132281 & 269.96 \\
					NVIDIA & N/A & 116 & 6425 & 69.84 \\
					Popkey & 12 & 84 & 11523 & 177.28 \\
					Tidal & 13 & 115 & 5339 & 58.03 \\
					Slack & 97 & 257 & 38411 & 185.56 \\
					Tofino Browser & 4 & 277 & 7365 & 31.21 \\
					SteelSeries & 1 & 4 & 364 & 182 \\
					Wire Messenger & 11 & 336 & 19306 & 70.20 \\
					\hline
					Average & 27 & 227 & 21803 & 104.95 \\
					\hline
				\end{tabularx}
				\caption{An automated scan of fifteen popular Electron applications}
				\label{tab:commitsBehind}
				\end{table}

\begin{figure}[ht] % Tasty graph
\centering
				\centering
				\includegraphics[width=88mm]{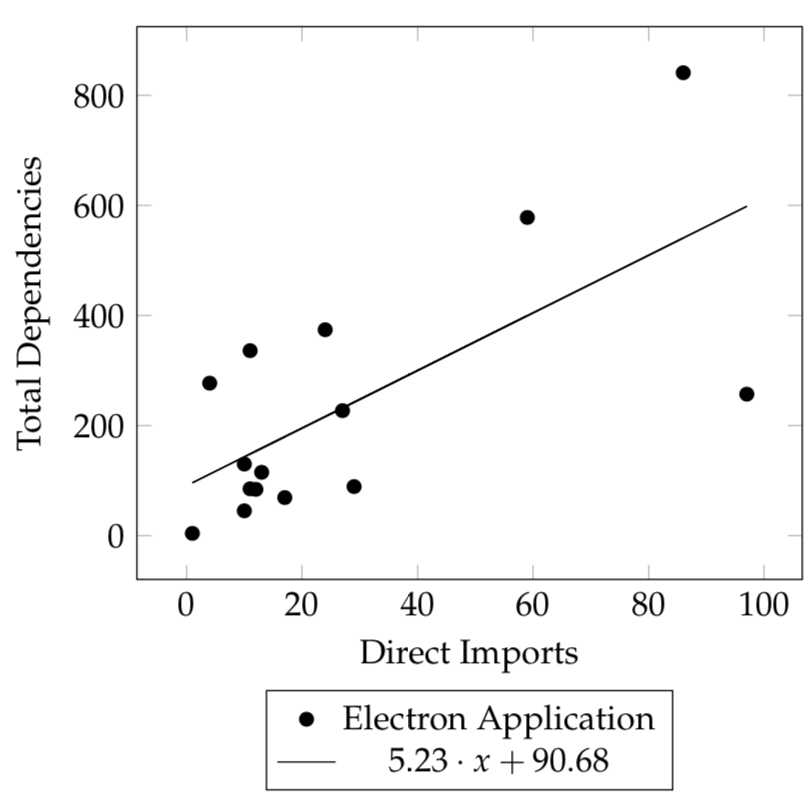}
				\caption{Graph demonstrating how an increase in first-hand imported modules directly causes a compound increase in external imported dependencies.}
				\label{graph:modules}
				\end{figure}

In addition, the number of application dependencies were compared to the number of modules explicitly included by the developers. The results (\autoref{graph:modules}) shows that a linear regression exists between the number of direct imports and the total number of dependencies. This demonstrates that a relatively low number of direct imports can result in mass inclusions of additional dependent modules, vastly increasing the code base with an approximate fivefold increase in total modules against primary imports.
In addition to the GitHub upstream checker, the same applications were audited against the Node Security Platform and the list of advisories that NSP maintain on npm modules. \autoref{tab:nspChecker} (previously shown in \autoref{nspCheck.js}) details the number of unique advisories found along with the highest CVSS score found.

\subsection{Update Injection and Malicious Modules}
By producing a malicious update package and placing it on a server which responded to URL requests from Electron applications, it was  possible to inject an application which had not taken basic security measures to ensure server authenticity. If the malicious update claimed a high version number (such as \texttt{v.50} when the current version is \texttt{v.0.8.1})  future updates would also be blocked as the application would opt out of downgrading its version number. Once injected (either through update hijacking or script running), it is possible to exfiltrate data through the execution of commands on the remote system.

\section{Discussion} \label{discussion}

\subsection{Current Remediations}
A number of suggestions have been made to combat the risk of uncontrolled JavaScript in web and client-side applications, however a lot of these struggle to find their place in the full-stack JavaScript development environment~\cite{kirda2006noxes}\cite{prokhorenko2016context}. For example \cite{whatYouInclude} recommends sandboxing remote code as well as manually importing remote includes by writing the required code directly into the main source.
While sandboxing has been a long discussed issue on the Electron GitHub repository and work was done by developers of the Brave Browser to reintroduce the Chrome Sandbox in their forked version of Electron \cite{electronSandboxGitHub}, there is still a lot of discussion and work to be done in order to merge these changes back into the upstream Electron repository \cite{sandboxProgress}. With regards to minimizing external includes, the issue lies with the Node.js model of programming where small modular includes are a major part of the language style. Whilst it would be possible to rewrite code from external modules directly into the source code, it is not the widely accepted or adopted method of Node.js application development. In addition to this, there are occasions where this is simply infeasible and could end up causing more harm than good --- for instance, with regards to cryptography and authentication modules.
By analyzing how other languages and frameworks handle code integrity, the immaturity of some elements of Node.js and Electron become clear. \cite{appInteg} poses multiple solutions for ensuring software integrity including the introduction of `file system integrity checkers', `code signing' and `visualization' in the event of attempted malicious code execution. Whilst some of these elements may be in the process of implementation, these measures simply haven't been effectively implemented in Electron.
There have been a number of methods proposed by which arbitrary code execution could be mitigated in both the context of secure application development and in application runtime monitoring and threat modelling. One such method is presented by \cite{synode} who proposed a tool that reduced the impact and effectiveness that vulnerable call sites (such as \texttt{eval} and \texttt{exec}) have on a system. Whilst this research study discusses \textit{``[their] runtime mechanism effectively prevents 100\% of the attacks''}, they also highlight that \textit{``developers who both use and maintain JavaScript libraries are reluctant to use analysis tools and are not always willing to fix their code''}. This reluctance to employ third-party security tools suggests the need to push greater security measures into the frameworks and runtimes themselves, rather than placing the onus on either the developer or the user to enforce good security practices at runtime.
Another solution to protect against malicious modules is based around runtime modification and surrounds itself with the idea of threat modeling. \cite{NodeSentry} investigates the process of creating \textit{``access control policies on interactions between libraries and their environment''}, giving modules only the access that they require based on a model of `least-privilege'. Whilst this is a potentially feasible implementation when dealing with server-side JavaScript, this policy enforcement can not be guaranteed when applied to Electron, as an attacker could have full control over the source code and execution of the application.

\subsection{Strength and Limitations of the Mayall Framework}

\subsubsection{Git Commit Divergence} \label{appScannerDiscussion}
As briefly mentioned in \autoref{appScanner.js}, whilst a notable divergence in commits from the master branch may not necessarily indicate the certainty of a security risk, it is still enough to raise concern. This disparity in module update mentality when compared to the urgency and immediacy of updates seen elsewhere (operating systems, etc.) may have been caused in part by this notion of version pinning, combined with the prolific lack of code maintenance --- an issue that was also noted by \cite{synode} in \autoref{devAttLit}.
Whilst version pinning will prevent unwanted change (be it breaking API changes or malicious trojan inserts), it may also simultaneously prevent security updates from being patched into a system. This was the case when modules in the test sample were found to contain known vulnerabilities when cross-referenced against the Node Security Platform Advisories List \cite{NSPAdvise}.
By modifying the \texttt{appScanner.js} tool to additionally search for key phrases such as `security', `urgent' or `vulnerability' within the commit messages or issues that are present between the current version and the version in use, it should be possible to create more meaningful data that developers will be able to respond to quickly. The primary risk with this method is centered around the list of search terms that commits and issues are compared against. Each developer and repository has their own type and there is a risk that tag names will be missed if they are stylized differently (for example a tag with the name `Type: Bug' instead of just `Bug'). If crucial tags are missed, then vulnerabilities could slip through the net, giving developers a false sense of security that their applications are secure.
The Node Security Platform is making strong steps towards providing a searchable database of known security vulnerabilities, however more work is required to ensure that module maintainers are submitting these vulnerabilities to the database as they patch, in order to provide a central queryable database for all developers.

\subsubsection{Update Process}
Evergreen applications are gaining greater traction as part of the software development life cycle. Evergreen here refers to the process of automatic self-updating and is adopted by widely used applications such as Chrome and Spotify as well as the majority of applications built on Electron. As a result of this well documented methodology, it is easy to configure an update server which applications can poll at regular intervals as part of this self-updating process. Consequently, it has also proved trivial for an attacker to configure a rogue server that mimics expected application endpoints if developers have not taken basic measures to protect the security of this process.

Whilst TLS encrypted communication and application code signing is not an issue caused directly by Electron, steps could be taken to enforce these more secure practices. For example, macOS applications can not be automatically updated unless a valid code signature is provided. Whilst this is a security feature present in macOS, this ideology could be transferred from the operating system paradigm into the cross-platform Electron Framework. By rejecting all application updates that do not have valid code signatures, developers will be forced to move towards code signing in order to release a popular and relevant application. Although this may seem like a drastic measure, Google are performing similar measures with regards to their ``Not Secure'' stance on web pages that do not employ HTTPS~\cite{notsecure}.
In addition to this, the introduction of code signing across the board may help to unify the fragmented update process currently in place. By calling for a redesign of the process, it may also be beneficial to reassess the packaging process so that the entire procedure can be made simpler for developers, potentially with the introduction of certificate acquisition scripts which are a practice currently recommended by \cite{certbot} with regards to TLS. This can also bring Linux \texttt{autoUpdating} inline with macOS and Windows, a platform which is currently unsupported by the Squirrel updater.

\subsubsection{Application Signing Against Runtime}
While the~\texttt{Program Files} directory in Windows requires administrative privileges to write to, the majority of Electron applications are installed in \texttt{AppData/Local} on Windows which does not require any special permissions. This places application source code in an unprotected environment and is free to be modified by any process. In order to protect the system from any malicious activity caused by a change in source code, code integrity checking mechanisms must be implemented at the execution stage of an application.
This is a process that is not only confined to Windows, as macOS also suffers the same issue. By running simple scripts with the same privilege of the current user, all Electron applications can be injected with malware, potentially without any knowledge of this from the user. If Electron were to implement rigid code signing policies, any malicious changes made to the source code of such applications would result in a non-execution of the application at launch.
Unfortunately, this change to the framework would create massive breaking changes. If the next version of Electron was to implement this type of code signing, there may not be anything stopping an attacker from replacing the Electron binary with a downgraded version which does not check for code signatures. As a result, application breaking code changes may be required in order to make downgrading a time inefficient attack for an adversary.

\subsubsection{Secure TLS Connections}
One major oversight was the lack of transport layer security implementations that allowed for malicious update injections due to the absence of cryptography on all connections to the update server. As Electron allowed updates to occur over HTTP, it was trivial to execute a man in the middle attack and perform arbitrary remote code execution on the target machine. If Electron were to prevent such connections from occurring, it would prove much more difficult to inject applications in this way. In addition to this, Electron should issue warnings to developers concerning the downloading of resources over HTTP and begin deprecation of this plaintext transport mechanism. At the time of writing, there are no less than 142 individual security advisories on the Node Security Platform for applications which download resources over HTTP, leaving themselves vulnerable to MITM attacks, potential code injection and arbitrary execution. This downloading of binary resources through insecure means opens applications up to the types of attack that have been demonstrated throughout this paper as an attacker will be able to replace the resource being downloaded with their own malicious versions.

\subsection{Repackaging binaries}
The practice of shipping JavaScript applications is becoming widespread, however the act of resigning a Node.js binary and shipping it as part of a wider product has not been properly observed or documented in the wild until it was discovered in an NVIDIA package by \cite{NVIDIANode}. By providing the Node.js binary to vast swathes of users, one of the main barriers to entry for JavaScript malware is removed. One potential explanation as to why JavaScript malware has only seen a slow uptick in recent years could be due to the lack of an available runtime to execute it against. As more applications are beginning to ship with Node.js included either through Electron or as a directly repackaged binary, it may be the case that a notable increase in JavaScript malware is observed. The ability to be able to produce truly cross-platform malware that can run on both servers and clients across all major operating systems could have potentially catastrophic outcomes.

\subsection{Developer Involvement with the Node Security Platform}
Although it may be easy for security researchers to pin the blame on the insecure practices of developers, this is neither helpful nor constructive. In order to remediate the risks associated with the use of known vulnerable modules, more awareness of the dangers and available scanning tools must be made available to developers. The npm package manager has the ability to issue warnings at the terminal prompt whenever vulnerable or deprecated modules are \texttt{npm install}ed, however these warnings do not flag up for every vulnerability listed in the NSP Advisories List.

It is proposed that npm issue a warning detailing the risks of installing these modules and have developers explicitly accept these risks before the installation continues. In addition to this, it may be beneficial to warn developers of out of date, dangerous modules at runtime, each time the application is tested or executed. This upfront alerting mechanism may influence developers to start taking measures to secure their applications before deployment.

\section{Conclusions}
\label{conclusion} % 885 words
In this manuscript a fully-functional framework is presented, demonstrating a technique to successfully inject user-installed applications with malicious Mayall modules.  The framework presented was made of two primary modules: the injector and the command and control server, as well as the vulnerability scanning algorithms. By using the Electron runtime as a standalone executable, the injector was able to execute on the remote host using the Electron and Node APIs bundled within the Electron binary. Furthermore, by providing the ability to pass \texttt{eval} and \texttt{exec} parameters to the payload from a remote server, execution is possible not only within the Electron and Node.js runtimes, but also from the native operating system shell.

The findings drawn from this manuscript signify the need for increased awareness and action surrounding JavaScript security. As Node.js inevitably continues to grow, the awareness of risks associated with running untrusted code need to grow alongside it. Unlike other popular programming languages (such as Python), Electron applications are not compiled directly into native executables for that operating system (in the same way py2exe would for Python) and are rather executed via the generic runtime environment that is distributed alongside the application. This introduces potential and previously unseen risks --- as demonstrated by NVIDIA's packaging and release of their rebranded and signed Node.js binary --- as it provides attackers with a new increasingly available toolset with inbuilt system interfacing capabilities. Moreover, this study demonstrated lack of accountability that an Electron binary held for its associated code base. The ability to entirely rewrite sections of an application without throwing alerts or warnings is a major risk that could be easily leveraged by an attacker under the appropriate circumstances.

\subsection{Future Work}
In order to be an extensible framework (similar to Powershell Empire and MetaSploit), the source code will be made available online for other researchers to contribute. By leveraging the module dependency nature of Node.js, we plan on  building extensions as standalone modules which are required by the core framework. 
Moreover, we plan to port popular penetration testing tools such as Mimikatz onto the Mayall framework.

%%%%%%%%%%%%%%%%%%%%%%%%%%%%%%%%%%%%%%%%%%
\vspace{6pt} 

%%%%%%%%%%%%%%%%%%%%%%%%%%%%%%%%%%%%%%%%%%
%% optional
%\supplementary{}

%%%%%%%%%%%%%%%%%%%%%%%%%%%%%%%%%%%%%%%%%%
%\acknowledgments{}

%%%%%%%%%%%%%%%%%%%%%%%%%%%%%%%%%%%%%%%%%%
\authorcontributions{Adam Rapley carried out the investigation, and wrote the first draft of the paper.  Colin McLean, Lynsay A. Shepherd, and Xavier Bellekens supervised the project, and wrote the paper.}
%%%%%%%%%%%%%%%%%%%%%%%%%%%%%%%%%%%%%%%%%%
\conflictofinterests{The authors declare no conflict of interest.} 

%%%%%%%%%%%%%%%%%%%%%%%%%%%%%%%%%%%%%%%%%%
%% optional
%\abbreviations{}

%%%%%%%%%%%%%%%%%%%%%%%%%%%%%%%%%%%%%%%%%%
%% optional
%\appendix
%\section{}
%%%%%%%%%%%%%%%%%%%%%%%%%%%%%%%%%%%%%%%%%%
\bibliographystyle{mdpi}
%=====================================
% References, variant B: external bibliography
%=====================================
\bibliography{references}
%%%%%%%%%%%%%%%%%%%%%%%%%%%%%%%%%%%%%%%%%%
\end{document}